\documentclass[aps,prc,twocolumn,superscriptaddress,showpacs,nofootinbib]{revtex4}

\usepackage{graphicx}
\usepackage{longtable}
\usepackage{dcolumn}
\usepackage{bm}
\usepackage{ulem}
\usepackage{color}
\usepackage{epsfig}

\begin{document}

\title{
Activation cross section measurement of the $^{14}$N(p,$\gamma$)$^{15}$O astrophysical key reaction
}
\author{Gy. Gy\"urky}%
\email{gyurky@atomki.hu}
\affiliation{Institute for Nuclear Research (ATOMKI), H-4001 Debrecen, Hungary}
\author{Z. Hal\'asz}%
\affiliation{Institute for Nuclear Research (ATOMKI), H-4001 Debrecen, Hungary}
\author{G.G.~Kiss}%
\affiliation{Institute for Nuclear Research (ATOMKI), H-4001 Debrecen, Hungary}
\author{T. Sz\"ucs}
\affiliation{Institute for Nuclear Research (ATOMKI), H-4001 Debrecen, Hungary}
\author{Zs. F\"ul\"op}%
\affiliation{Institute for Nuclear Research (ATOMKI), H-4001 Debrecen, Hungary}

\date{\today}

\begin{abstract}
\begin{description}

\item[Background]

$^{14}$N(p,$\gamma$)$^{15}$O is one of the key reactions of nuclear astrophysics playing a role in various stellar processes and influencing energy generation of stars, stellar evolution and nucleosynthesis. For a reliable reaction rate calculation the low energy cross section of $^{14}$N(p,$\gamma$)$^{15}$O must be known with high accuracy. Owing to the unmeasurable low cross sections, theoretical calculations are unavoidable. 

\item[Purpose]

High precision experimental cross section data are needed in a wide energy range in order to provide the necessary basis for low energy extrapolations. In the present work the total $^{14}$N(p,$\gamma$)$^{15}$O cross section was measured with a method complementary to the available data sets. 

\item[Method]

The cross section was measured with activation, based on the detection of the annihilation radiation following the $\beta^+$-decay of the reaction product $^{15}$O. This method, which provides directly the astrophysically important total cross section, was never used for the $^{14}$N(p,$\gamma$)$^{15}$O cross section measurement in the studied energy range.

\item[Results]

The non-resonant cross section was measured between 550\,keV and 1400\,keV center-of-mass energies with total uncertainty of about 10\,\%. The results were compared with literature data using an R-matrix analysis. It is found that the cross sections measured in this work are in acceptable agreement with the two recent measurements only if the weak transitions -- not measured in those works -- are included. 

\item[Conclusions]

The present data set, being largely independent from the other available data, can be used to constrain the extrapolated cross sections to astrophysical energies and helps to make the astrophysical model calculations more reliable.

\end{description}
\end{abstract}

\pacs{26.20.Cd,25.40.Lw
}

\maketitle


Hydrogen burning is perhaps the most important source of energy in our universe as all stars start their lives by fusing hydrogen into helium \cite{Adelberger2011}. Stars less massive than about 1.3 solar masses burn hydrogen dominantly through the pp-chain reactions, while more massive stars utilize the CNO cycle, a catalytic cycle of reactions on C, N and O isotopes \cite{Wiescher2018}. Our Sun belongs to the first group, roughly 98\,\% of its energy is thus generated by the pp-chains. The remaining two percent, however, is still highly important as the exact share of the CNO cycle in the energy budget is correlated with the metallicity of the Sun \cite{Haxton2008}. The heavy element content of the solar core is an extensively studied quantity as contradicting values are derived from the standard solar model using photospheric abundances and from helioseismic observations \cite{Villante2019}. The recent observation of neutrinos from the CNO cycle by the Borexino detector \cite{Borexino2020} allows the direct measure of the CNO reaction rate in the Sun. In order to infer the solar metallicity from the measured CNO neutrino flux, the rates of nuclear reactions in the cycle must be known with good accuracy \cite{Orebigann2021}. The importance of the CNO cycle in other astrophysical scenarios is discussed in detail e.g. in Ref.\,\cite{Wiescher2010}.

The slowest reaction of the CNO cycle is $^{14}$N(p,$\gamma$)$^{15}$O determining thus the rate of the whole cycle. Its relevant energy range at solar temperature is between about 20 and 35\,keV where the extremely low cross section prevents any direct measurement with the presently available experimental techniques. The reaction rate is thus determined from theoretical cross sections which are often obtained from the R-matrix theory. For an R-matrix extrapolation, measured cross sections at higher energies are needed. 

Because of its astrophysical importance, the $^{14}$N(p,$\gamma$)$^{15}$O cross section was measured many times. The first measurements in the 1950s and 1960s are not included in modern compilations due mainly to the lack of sufficient experimental information (for references and discussions see Ref.\,\cite{Adelberger1998}). In 1987, Schr\"oder \textit{et al.} \cite{Schroder1987} measured the cross section in a wide energy range, but its result had to be corrected later for experimental errors \cite{Adelberger2011}. More recent experiments concentrated mainly on the lowest measurable energies and were thus carried out in relatively narrow energy ranges \cite{Formicola2004,Lemut2006,Bemmerer2006,Runkle2005,Imbriani2005,Marta2011}. However, due to the relatively complicated structure and decay scheme of $^{15}$O, for a reliable extrapolation the cross section must also be known at higher energies including both the direct capture component and the parameters of wide and narrow resonances. Recently the results of two experiments in wide energy ranges became available \cite{Li2016,Wagner2018}. The zero energy extrapolated $S$-factor values\footnote{The astrophysical $S$-factor as a function of interaction energy $E$ is defined as $S(E$)=$\sigma(E)\cdot E \cdot e^{2\pi\eta}$ where $\eta$ is the so-called Sommerfeld parameter. It is used to remove the strong energy dependence of the cross section $\sigma(E)$ due to the Coulomb barrier penetration \cite{Iliadis2007}.}  based on these experiments agree very well for the transition to the 6.79\,MeV excited state but unfortunately the ground state cross section extrapolations differ by as much as a factor of two and bear high uncertainty. This means that the astrophysical reaction rate at solar temperatures still cannot be calculated from the available cross sections at the required precision \cite{Orebigann2021} and nuclear physics input remains one of the most important sources of uncertainty of solar models. Further experiments are thus clearly needed and independent approaches can be very useful in order to check the reliability of the available, but often contradicting data. 

In the present work the $^{14}$N(p,$\gamma$)$^{15}$O non-resonant cross section was measured in the center-of-mass energy range between 550 and 1400 keV using the activation method \cite{Gyurky2019}, which was not applied in any recent study of this reaction. The activation method is possible since the product of the reaction, $^{15}$O is radioactive, decays by positron emission to $^{15}$N with a half-life of 122.24\,$\pm$\,0.16\,s \cite{Ajzenberg1991}. The decay can be observed by detecting the 511\,keV $\gamma$-radiation following the positron annihilation. As the activation method is based on the determination of the number of reaction products, it gives directly the total cross section and cannot provide partial cross sections for the various transitions (like, e.g. the above mentioned two most important transitions to the ground and to the 6.79\,MeV exited states). The measured total cross sections, however, are largely independent from the ones based on in-beam $\gamma$-spectroscopy and can be used to constrain R-matrix fits. Moreover, this method is free from some uncertainties burdening the in-beam experiments such as angular distribution effects or the true coincidence summing. 

The present work is the continuation of our previous experiment where the strengths of two narrow resonances in $^{14}$N(p,$\gamma$)$^{15}$O were measured with activation \cite{Gyurky2019b}. As the experimental technique is the same as in that work, here only the most important features are summarized. 

Solid state TiN target were used for the experiments which were prepared by reactive sputtering onto thick Ta backings. Targets with three different thicknesses (100, 200 and 300\,nm) were used. The absolute thicknesses and the Ti:N atomic ratios were measured with four independent techniques: Rutherford Backscattering Spectrometry, Proton Induced X-ray Emission, Secondary Neutral Mass Spectrometry and Nuclear Resonant Reaction Analysis \cite{Gyurky2019b,Gyurky2017b}. Based on these measurements, the number of target atoms, as the quantity required for the cross section calculation, was determined to a precision of 5\,\%.

The targets were irradiated by proton beams provided by the Tandetron accelerator of Atomki in Debrecen, Hungary \cite{Rajta2018,Biri2021}. The typical beam intensity was about 5\,$\mu$A on target. The studied energy range between $E_{\rm p}$\,=\,600 and 1500\,keV was covered with 50-100\,keV energy steps. Owing to the short half-life of $^{15}$O, the measurement was carried out with a cyclic activation. The beam bombarded the targets for five minutes which was followed by a 10 or 20 minute beam-off period used for decay counting. Depending on the cross section, the cycle was repeated minimum four and maximum twenty-two times. 

The 511\,keV positron-annihilation $\gamma$-radiation following the $^{15}$O decay was measured with a 100\,\% relative efficiency HPGe detector placed behind the target chamber at a distance of roughly 1\,cm from the target. The absolute detection efficiency of the detector in the non-trivial geometry of the positron annihilation (the positrons emitted towards the vacuum chamber annihilate at different positions) was measured using the procedure described in detail in Ref.\,\cite{Gyurky2019b}. 

\begin{figure}
\includegraphics[width=\columnwidth]{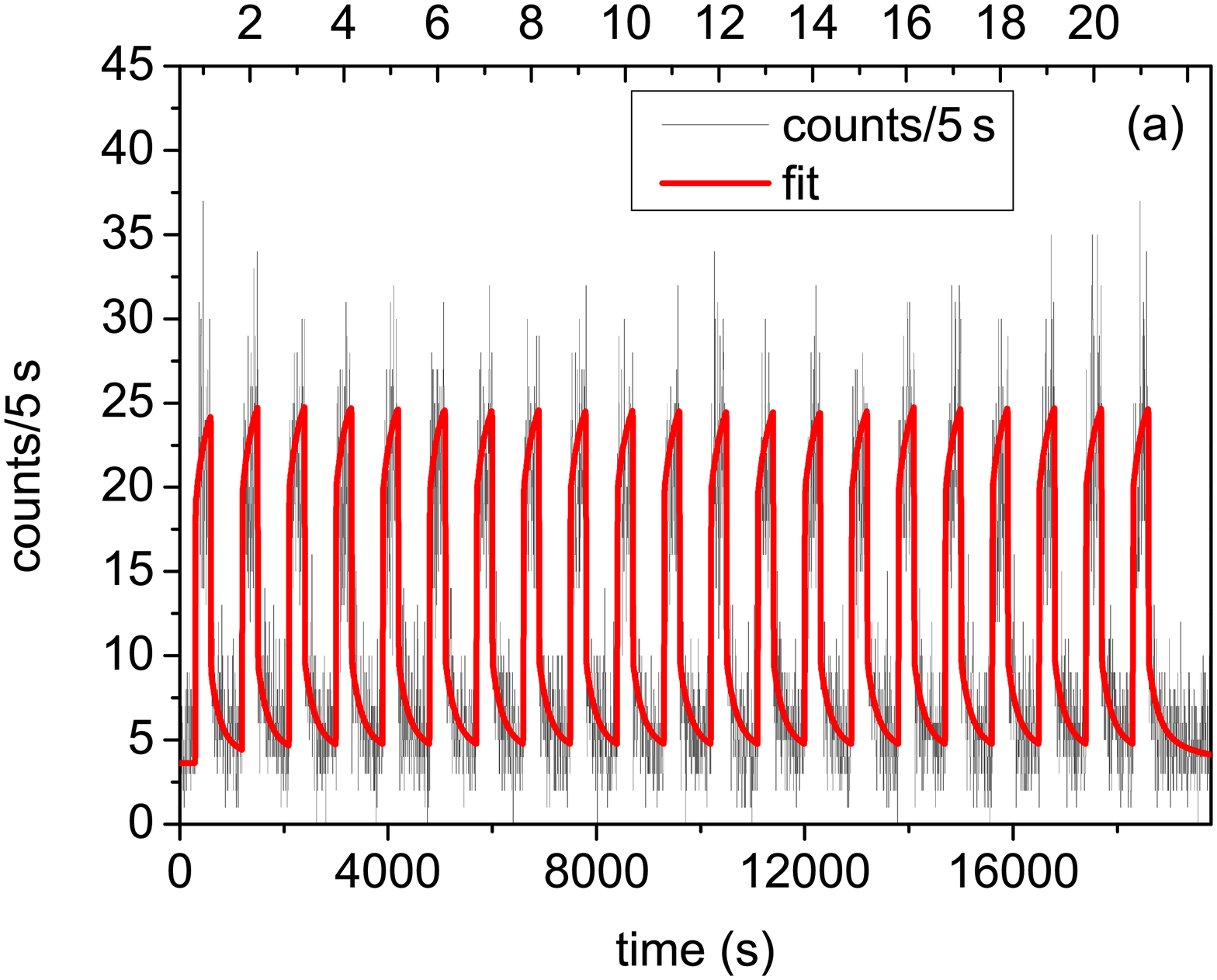}\\
\includegraphics[width=\columnwidth]{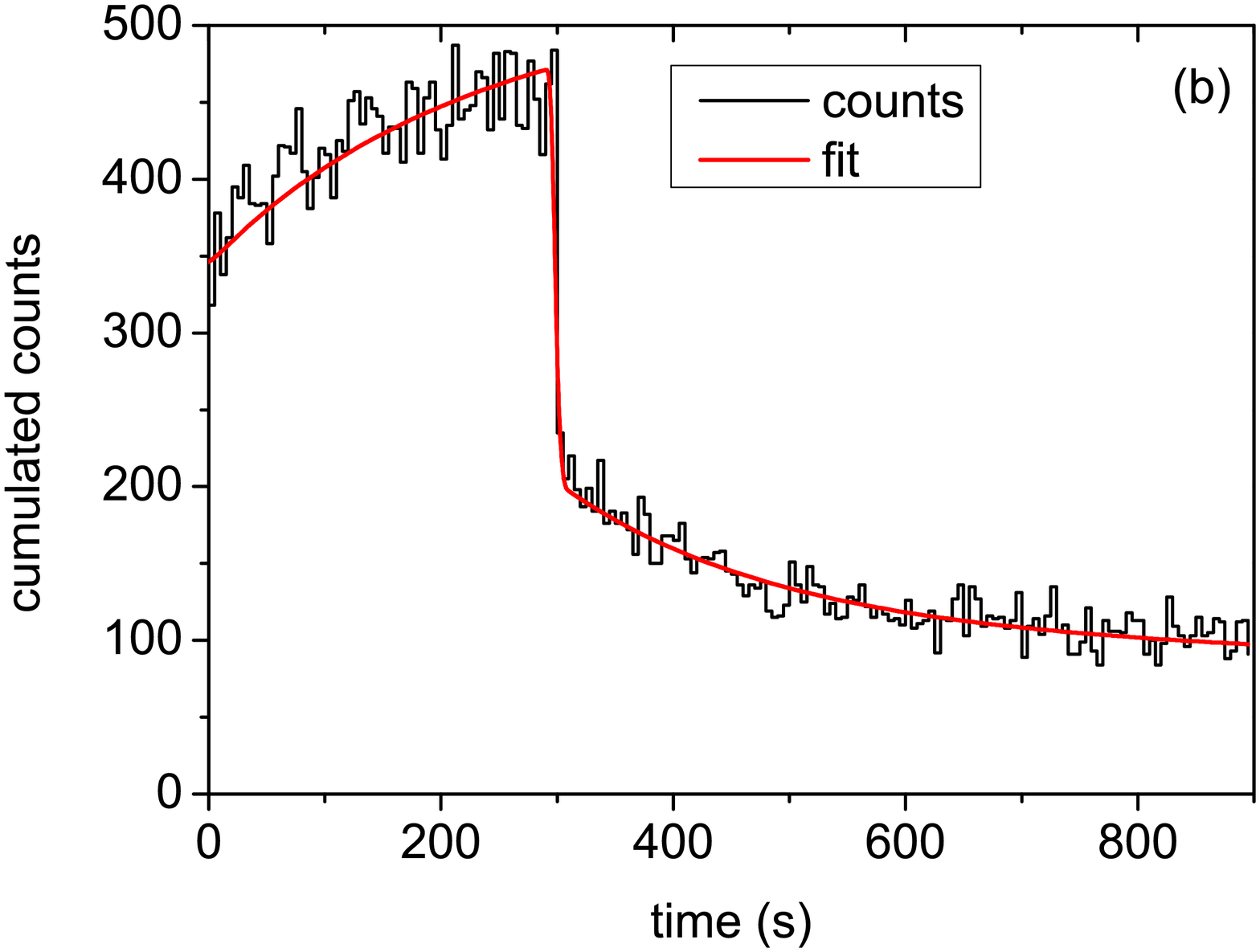}
\caption{\label{fig:decay} Upper panel: number of detected 511\,keV $\gamma$-rays in the five second intervals as a function of time in the case of an activation of 21 activation cycles (the upper axis shows the cycle number). Lower panel: The cumulative number of detected events summing up the 21 cycles. The fit of the  decay curves (shown by smooth red lines) were used to obtain the cross section.}
\end{figure}

The counts in the 511\,keV peak were recorded every 5 seconds during the cyclic activation process. As an example, the upper panel of Fig.\,\ref{fig:decay} shows the number of detected events as a function of time in the case of a measurement with 21 cycles. In order to better see the decay of the reaction products, the lower panel shows the same run, but summing up the 21 cycles. The decay curves were fitted using the well known half-life of $^{15}$O and from the fit the number of created $^{15}$O isotopes and hence the cross section could be determined. 

Naturally, the 511\,keV $\gamma$-radiation does not contain information about its origin, any positron emitter radioactive isotope or the pair production of higher energy $\gamma$-photons can result in such a $\gamma$-ray. It is thus crucial to identify that the detected annihilation radiation is from the decay of $^{15}$O. The half-life analysis of the decay curve can be used for this purpose. Two parasitic reactions were identified which contributed to the detected 511\,keV events. The $^{12}$C(p,$\gamma$)$^{13}$N reaction, induced on the carbon built-up of the target surface, has a strong wide resonance at $E_{\rm p}\approx\,$420\,keV. The 9.965 minute half-life of $^{13}$N is long enough to distinguish its decay from that of $^{15}$O. The increasing cross section of $^{12}$C(p,$\gamma$)$^{13}$N towards lower energies, along with the dropping $^{14}$N(p,$\gamma$)$^{15}$O cross section limited our measurement to proton energies of 600\,keV and above. At the upper part of the studied energy range, the cross section of the $^{46}$Ti(p,$\gamma$)$^{47}$V reaction starts to be significant. This reaction is unavoidable if TiN targets are used. The half-life of the positron emitter $^{47}$V is rather long, 32.6\,min, can be easily distinguished from $^{15}$O. However, the increasing background caused by this parasitic reaction prevented us from measuring the $^{14}$N(p,$\gamma$)$^{15}$O cross section above 1500\,keV proton energy.

\begin{table}
\caption{\label{tab:results} Measured cross section of the $^{14}$N(p,$\gamma$)$^{15}$O reactions. For details see text.}
\begin{tabular*}{\columnwidth}{@{\hskip\tabcolsep\extracolsep\fill}rr@{\hspace{-5mm}}c@{\hspace{-5mm}}llll}
\hline
$E_{\rm beam}$ & \multicolumn{3}{c}{$E^{\rm eff}_{\rm c.m.}$} & $\sigma$ & $\Delta\sigma^{\rm stat.}$ & $\Delta\sigma^{\rm total}$ \\
 \cline{5-7}
{[keV]}				&	 \multicolumn{3}{c}{[keV]}		& \multicolumn{3}{c}{[$\mu$barn]}		\\
\hline
599.6	&	549.6	&	$\pm$	&	8.5	&	0.398	&	0.041	&	0.049	\\
649.3	&	596.5	&	$\pm$	&	8.0	&	0.469	&	0.035	&	0.048	\\
699.3	&	643.6	&	$\pm$	&	7.7	&	0.557	&	0.024	&	0.046	\\
749.3	&	690.6	&	$\pm$	&	7.4	&	0.731	&	0.061	&	0.080	\\
799.2	&	737.6	&	$\pm$	&	7.1	&	0.816	&	0.038	&	0.069	\\
849.2	&	784.5	&	$\pm$	&	6.8	&	1.10	&	0.07	&	0.10	\\
899.2	&	831.4	&	$\pm$	&	6.6	&	1.21	&	0.06	&	0.10	\\
949.2	&	878.3	&	$\pm$	&	6.4	&	1.39	&	0.07	&	0.12	\\
999.1	&	925.1	&	$\pm$	&	6.3	&	1.90	&	0.07	&	0.15	\\
1147.1	&	1063.7	&	$\pm$	&	5.9	&	2.02	&	0.05	&	0.15	\\
1199.0	&	1112.3	&	$\pm$	&	5.8	&	2.26	&	0.07	&	0.17	\\
1299.0	&	1205.8	&	$\pm$	&	5.5	&	2.82	&	0.05	&	0.21	\\
1398.9	&	1299.4	&	$\pm$	&	5.3	&	3.62	&	0.05	&	0.26	\\
1498.8	&	1393.1	&	$\pm$	&	4.9	&	4.55	&	0.14	&	0.35	\\
\hline
\end{tabular*}
\end{table}

Table \ref{tab:results} shows the measured cross sections. The first column contains the primary proton beam energies which are known to a precision of better than 1\,keV \cite{Csedreki2020}. In the second column the effective center-of-mass energies are listed which were calculated by taking into account the energy loss of the beam in the TiN layers. Since this energy loss is relatively low (between about 5 and 20\,keV center-of-mass energy) and the cross section changes by no more than a few percent in such an energy range, the effective energies are the arithmetic mean of the entrance and exit energies. The quoted uncertainties of $E^{\rm eff}_{\rm c.m.}$  take into account the target thickness, stopping power and primary beam energy uncertainties. The measured cross sections are shown in the third column while the last two columns contain the statistical and total cross section uncertainties, respectively. The total uncertainties were obtained by adding in quadrature the following systematic uncertainties to the statistical one: number of target atoms (5\,\%), detector efficiency (4\,\%) and beam current integration  (3\,\%). Other sources of systematic uncertainty (like e.g. the $^{15}$O decay parameters) are well below 1\,\% and thus neglected. 

The comparison of the present results with literature data is not straightforward as no total (i.e. including all transitions), angle integrated cross sections are available in the studied energy range. The only experiment which provided total cross section was carried out by the LUNA collaboration, however, at much lower energies \cite{Bemmerer2006}. 

\begin{figure}
\includegraphics[width=\columnwidth]{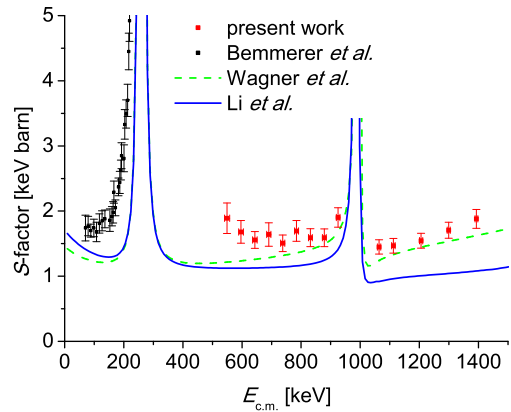}
\caption{\label{fig:results} Measured cross section of the $^{14}$N(p,$\gamma$)$^{15}$O reaction in the form of astrophysical $S$-factor. Besides the present results, the experimental data of Ref.\,\cite{Bemmerer2006} are shown as well as the sum of the cross sections of the two dominant transitions obtained from the R-matrix fits of Refs.\,\cite{Li2016,Wagner2018}}
\end{figure}

In the two recent experiments, the two strongest transitions (to the ground state and to the 6.79\,MeV excited state of $^{15}$O, including their angular distributions) were measured in an energy range overlapping with the present one \cite{Li2016,Wagner2018}. An R-matrix fit to those data was carried out in both works which allows the calculation of the summed cross section of the two studied transitions. The comparison of our total cross section with those results is shown in Fig.\,\ref{fig:results} in the form of $S$-factor. The plotted lines were obtained using the AZURE R-matrix code \cite{AZURE} with the parameters listed in the original publications \cite{Li2016,Wagner2018} and provided by the authors \cite{Wagner_private}. Our data are significantly higher than those of Li \textit{et al.} \cite{Li2016} and somewhat higher than (but compatible within two standard deviations with) the results of Wagner \textit{et al.} \cite{Wagner2018}. This deviation is not surprising as weaker transitions -- not measured in those works -- also contribute to the total cross section in the studied energy range and those were not included in the shown R-matrix calculations. The effect of the missing transitions is evident also around the $E_{\rm c.m.}$\,=\,259\,keV resonance. In the figure, the total cross sections of Bemmerer \textit{et al.} \cite{Bemmerer2006} are also plotted which shows that both R-matrix curves based solely on the two strongest transitions result in a much narrower resonance, in contradiction with the measurement. 

\begin{figure}
\includegraphics[width=\columnwidth]{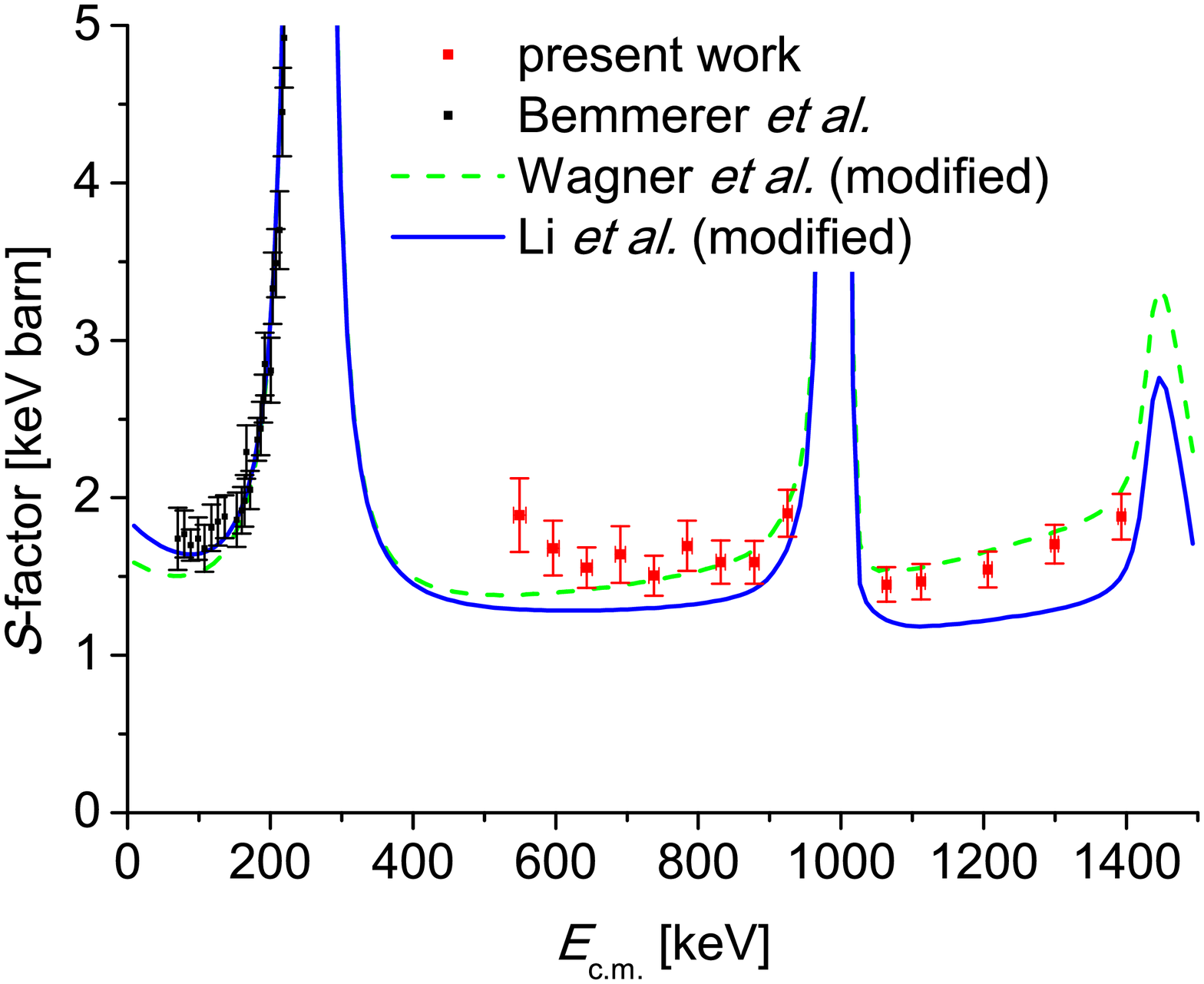}
\caption{\label{fig:results_mod} Same as Fig.\,\ref{fig:results}, but the R-matrix calculations contain also the transitions not measured in Refs.\,\cite{Li2016,Wagner2018}. The partial $S$-factors for these transitions are taken from the work of Imbriani \textit{et al.} \cite{Imbriani2005}.}
\end{figure}

In order to remedy the problem of missing transitions, the $S$-factors of the two dominant transitions measured and fitted by Li \textit{et al.} and  Wagner \textit{et al.} were complemented by including the contributions of weaker transitions, namely the ones to the 5.18\,MeV, 5.24\,MeV and 6.17\,MeV states. These latter $S$-factor values were taken from the analysis of Imbriani \textit{et al.} \cite{Imbriani2005}. The results are shown in Fig.\,\ref{fig:results_mod}. The inclusion of the weaker transitions resulted in a wider $E_{\rm c.m.}$\,=\,259\,keV resonance, in good agreement with the experimental results of Bemmerer \textit{et al.} \cite{Bemmerer2006}, as well as an increased direct capture cross section in the energy region of the present experiment. In this region the results of Wagner \textit{et al.} \cite{Wagner2018} are in relatively good agreement with the present data.
The data of Li \textit{et al.} \cite{Li2016} are always below our values and are compatible only at the level of two standard deviations. 

In summary, the total cross section of the $^{14}$N(p,$\gamma$)$^{15}$O reaction was measured in the present work using the activation method in the energy range between $E_{\rm c.m.}$\,=\,550\,keV and 1400\,keV. Based on a simple R-matrix calculation, the results are compared with two recent measurements of partial cross sections. Although the most important astrophysical quantity is the zero energy extrapolated $S$-factor, we do not intend to provide such a value based solely on our data, as the total cross section does not provide information on the contribution of various transitions. On the other hand, activation is a largely different method from the typically applied in-beam $\gamma$-spectroscopy. Therefore, our data provide an independent anchor point for R-matrix fits and comparison possibility with other data sets. The comparison with the results of the two recent measurements of Wagner \textit{et al.} and Li \textit{et al.} \cite{Wagner2018,Li2016} shows that reasonable agreement is found only if the weak transitions -- not measured in those works -- are included. This indicates that further experiments are needed in this energy range -- and in general in an energy range as wide as possible from the lowest measurable energies to the several MeV range -- including all the relevant transitions. This is necessary in order to provide high precision nuclear input data for the modern astrophysical models. 

\begin{acknowledgments}
This work was supported by the European COST action "ChETEC" (CA16117), by NKFIH grants No. NN128072 and K134197 and by the \'UNKP-21-5-DE-479 New National Excellence Programs of the Ministry of Human Capacities of Hungary. The financial support of the Hungarian Academy of Sciences (Infrastructure grants), and the Hungarian Government, Economic Development and Innovation Operational Programme (GINOP-2.3.3–15-2016-00005) grant, co-funded by the EU, is also acknowledged. T. Sz\"ucs acknowledges support from the Bolyai research fellowship of
the Hungarian Academy of Sciences. The authors thank I. Rajta, I. Vajda and the operators of the Tandetron accelerator for providing excellent beams and working conditions. We also thank A. Cs\'\i k, R. Husz\'ank, Zs. T\"or\"ok, L. Wagner and M. G. Kohan for their contribution to the early phase of the experiments. 
\end{acknowledgments}

\bibliographystyle{apsrev}

\end{document}